\documentclass[preprint*]{JHEP3}
\usepackage{amsmath} % 10pt is ignored!

\usepackage{epsfig,multicol,bbm}

\newcommand\fverb{\setbox\pippobox=\hbox\bgroup\verb}
\newcommand\fverbdo{\egroup\medskip\noindent%
            \fbox{\unhbox\pippobox}\ }
\newcommand\fverbit{\egroup\item[\fbox{\unhbox\pippobox}]}
\newbox\pippobox
\def\c{\mathfrak{c}}
\def\p{\mathfrak{p}}

\title{Inflationary universe in loop quantum cosmology}

\author{Xin Zhang\\
Kavli Institute for Theoretical Physics China, Institute of
Theoretical Physics, Chinese Academy of Sciences (KITPC/ITP-CAS),
P.O.Box 2735, Beijing 100080, People's Republic of
China\\
    E-mail: \email{zhangxin@itp.ac.cn}}
\author{Yi Ling\\
Center for Gravity and Relativistic Astrophysics,
Nanchang University, Nanchang 330047, People's Republic of China\\
    E-mail: \email{yling@ncu.edu.cn}}

\abstract{Loop quantum cosmology provides a nice solution of
avoiding the big bang singularity through a big bounce mechanism in
the high energy region. In loop quantum cosmology an inflationary
universe is emergent after the big bounce, no matter what matter
component is filled in the universe. A super-inflation phase without
phantom matter will appear in a certain way in the initial stage
after the bounce; then the universe will undergo a normal inflation
stage. We discuss the condition of inflation in detail in this
framework. Also, for slow-roll inflation, we expect the imprint from
the effects of the loop quantum cosmology should be left in the
primordial perturbation power spectrum. However, we show that this
imprint is too weak to be observed.}

\begin{document}

%\maketitle  IS IGNORED %%%%%%%%%%%

\section{Introduction}

The inflation paradigm provides an exquisite explanation for some
severe problems of the cosmological standard model by positing an
epoch of accelerated expansion in the early universe
\cite{Guth,Starobinsky:infl}. This accelerated period of expansion
also generates superhorizon fluctuations and thus predicts an almost
scale-invariant density perturbation power spectrum, which has
received strong observational support from the measurement of the
temperature fluctuation in the cosmic microwave background (CMB)
radiation \cite{cobe,wmap1,wmap3}. Conceptually, however, the
inflationary scenario is incomplete due to the existence of the big
bang singularity \cite{Borde:2001nh}. Einstein's classical theory of
general relativity (GR) breaks down near such a singularity since
quantum effects are expected to be important at very high energies
in the early universe. So, the classical theory of GR has to be
replaced by some theoretical framework of quantum gravity which
should remain well defined even at very high curvatures.

Loop quantum gravity (LQG) is a leading nonperturbative background
independent approach to quantizing gravity \cite{LQG}. The
underlying geometry in LQG is discrete and the continuum spacetime
can be obtained from the quantum geometry in a large eigenvalue
limit. Loop quantum cosmology (LQC) focuses on symmetry reduced
models (with homogeneous and isotropic space) but inherits
quantization scheme and techniques from LQG \cite{LQC}.
Investigations of LQC has led to important insights on the
resolution of singularities in various situations
\cite{solsingu1,solsingu2,solsingu3}. Within the framework of LQC,
some long-standing issues concerning the quantum nature of the
big-bang are resolved in the context of homogeneous and isotropic
universe with a scalar field. Using extensive analytical and
numerical methods, the analysis of the evolution of the
semiclassical states for a spatially flat universe has shown that
the universe has a pre-big-bang branch, joined deterministically to
the post-big-bang branch by a quantum bounce in the deep Planck
regime through the LQC evolution \cite{quantum1,quantum2,quantum3}.
Thanks to the nonperturbative background independent methods of LQC,
the idea of the nonsingular bounce can be realized in a natural
fashion.

An important feature of LQC is that the underlying dynamics is
governed by a discrete quantum difference equation of quantum
geometry. An effective description of quantum dynamics, however, can
be obtained by applying geometric methods to quantum mechanics,
where the Hilbert space is treated as an infinite dimensional phase
space which has a structure of a fiber bundle. Using semiclassical
states one can construct an effective Hamiltonian description on a
continuum spacetime which incorporates the leading quantum
corrections to the classical dynamics and has been shown to be a
very good approximation to the quantum dynamics
\cite{quantum2,quantum3,semiclass}. Intriguingly, one can obtain an
modified Friedmann equation from the effective Hamiltonian
constraint, which can be used to investigate the role of
nonperturbative quantum correction conveniently. It is remarkable
that the quantum geometric effects lead to a $\rho^2$ modification
to the Friedmann equation at the scales when $\rho$ becomes
comparable to a critical density $\rho_{\rm crit}$ which is close to
the Planck density ($\rho_{\rm crit}\approx 0.82\rho_{\rm Pl}$)
\cite{quantum2,quantum3,dualRS,geom,bounce}. The modified term in
the Friedmann equation is negative definite, which implies a bounce
when the energy density hits the critical value; this feature
resolves the classical big-bang singularity problem and is in
accordance with the result from the quantum evolution in LQC.

Inflation begins after the big bounce when the quantum geometric
effects are dominant. Some questions naturally arise: What influence
does the quantum geometry make on the inflation? Can any matter
(with arbitrary equation of state) lead to inflation under the
consideration of LQC? What conditions are needed for inflation in
the effective LQC? Can the LQC effects provide sufficient number of
$e$-foldings for inflation? Since inflation happens at the very
early time, the effects of quantum gravity should in principle leave
an imprint on the primordial spectrum of perturbations, because the
wavelengths of perturbations emerged from short distances in early
stages of inflation are stretched to the cosmic scales observable
today by the rapid expansion during inflation. So, relevant
questions are: How the LQC effects affect the primordial power
spectra? Can LQC leave imprint on the CMB sky? Can we observe it?

This work seeks to answer these questions using the effective theory
of LQC. This paper is organized as follows. In the next section we
briefly review the effective theory of LQC and describe the quantum
big-bounce of the universe generated by loop quantum dynamics. In
section 3, we discuss some issues of inflation using the modified
Friedmann equation in the effective theory of LQC. We analyze the
condition of inflation in detail taking the LQC effects into
account; we discuss the $e$-folding number relevant to the quantum
gravity effects; we calculate the primordial power spectra of
density perturbations and gravitational waves, spectral indices and
tensor-to-scalar ratio in the slow-roll inflation when considering
the influences of the LQC; and we also analyze the observational
constraints on the LQC imprint on the CMB sky. We give conclusions
in section 4.

\section{Effective dynamics in loop quantum cosmology and big bounce of the universe}

LQG is a canonical quantization of gravity based upon
Ashtekar-Barbero connection variables. The phase space of classical
GR in LQG is spanned by SU(2) connection $A_a^i$ and the triad
$E_i^a$ on a 3-manifold $\cal{M}$ (labels $a$ and $i$ denote space
and internal indices respectively), which are two conjugate
variables encoding curvature and spatial geometry, respectively.
Likewise, LQC is a canonical quantization of homogeneous spacetimes
based upon techniques used in LQG. In LQC, due to the symmetries of
the homogeneous and isotropic spacetime, the phase space structure
is simplified, i.e., the connection is determined by a single
quantity labeled $\c$ and likewise the triad is determined by a
parameter $\p$. The variables $\c$ and $\p$ are canonically
conjugate with Poisson bracket $\{\c, \p\}=\gamma\kappa/3$, where
$\kappa=8\pi G$ ($G$ is the Newton's gravitational constant) and
$\gamma$ is the dimensionless Barbero-Immirzi parameter which is set
to be $\gamma\approx 0.2375$ by the black hole thermodynamics in LQG
\cite{BH}. For the spatially flat model of cosmology, the new
variables have the relations with the metric components of the
Friedmann-Robertson-Walker (FRW) universe as
\begin{equation}
\c=\gamma\dot{a},~~~~\p=a^2,\label{newvars}
\end{equation}
where $a$ is the scale factor of the universe. Classically in terms
of the connection-triads variables the Hamiltonian constraint is
given by
\begin{equation}
{\cal H}_{\rm cl}=-{3\sqrt{\p}\over \kappa\gamma^2}\c^2+{\cal
H}_{\rm M},
\end{equation}
where ${\cal H}_{\rm M}$ is the matter Hamiltonian.

The elementary variables used for quantization in LQC are the triads
and holonomies of the connection. The holonomy over an edge of a
loop is defined as $h_i(\mu)=\cos (\mu\c/2)+2\sin (\mu\c/2)\tau_i$,
where $\tau_i$ is related to Pauli spin matrices as
$\tau_i=-i\sigma_i/2$ and dimensionless $\mu$ is related to the
physical length of the edge over which holonomy is evaluated (note
that $\mu$ is also the eigenvalue of the triad operator $\hat{\p}$).
In the Hamiltonian formulation for homogeneous and isotropic
spacetime, the dynamical equations can be determined completely by
the Hamiltonian constraint. Under quantization, the Hamiltonian
constraint gets promoted to an operator and the quantum wave
functions are annihilated by the operator of the Hamiltonian
constraint. In LQC, it is expected that modifications due to LQC
effects will appear in the Hamiltonian constraint, and from the
modified Hamiltonian constraint the effective Friedmann constraint
will be derived. In quantization the Hamiltonian constraint operator
is obtained by promoting the holonomies and the triads to the
corresponding operators. Consequently, this leads to a discrete
quantum difference equation, which indicates that the underlying
geometry in LQC is discrete \cite{solsingu1,solsingu2}.
Interestingly, the solutions of this difference equation are
nonsingular.

So far we see that the underlying dynamics in LQC is governed by a
discrete quantum difference equation in quantum geometry. However,
an effective Hamiltonian description on a continuum spacetime can be
constructed by using semiclassical states, which has been shown to
very well approximate the quantum dynamics \cite{quantum2,quantum3}.
This analysis reveals that on backward evolution of our expanding
phase of the universe, the universe bounces at a critical density
(near the big bang singularity) into a contracting branch
\cite{quantum1,bounce}. Thus the classical singular problem can be
successfully overcome within the context of LQC by a nonsingular
bounce. In addition, the effective equations for the modified
Friedmann dynamics can be derived from the effective Hamiltonian
constraint with loop quantum modifications, which can be used to
investigate the role of nonperturbative quantum corrections. An
important feature for the modified dynamics is that a $\rho^2$ term
which is relevant in the high energy regime is included in the
classical Friedmann equation. The modified term is negative definite
implying a bounce when the energy density reaches a critical value
on the order of the Planck density.

The effective Hamiltonian constraint, to leading order, is given by
\cite{semiclass}
\begin{equation}
{\cal H}_{\rm eff}=-{3\over \kappa\gamma^2\bar{\mu}^2}a
\sin^2(\bar{\mu}\c)+{\cal H}_{\rm M},
\end{equation}
where $\bar{\mu}$ is the kinematical length of the edge of a square
loop which has the area given by the minimum eigenvalue of the area
operator in LQG; the area is ${\cal A}=\bar{\mu}^2 a^2=\alpha l_{\rm
Pl}^2$, where $\alpha$ is of the order unity and $l_{\rm
pl}=\sqrt{G}$ is the Planck length. The modified Friedmann equation
can then be derived by using the Hamilton's equation for $\p$,
\begin{equation}
\dot{\p}=\{\p, {\cal H}_{\rm eff}\}=-{\kappa\gamma\over 3}{\partial
{\cal H}_{\rm eff}\over \partial
\c}={2a\over\gamma\bar{\mu}}\sin(\bar{\mu}\c)\cos(\bar{\mu}\c),
\end{equation}
which combined with Eq. (\ref{newvars}) yields the rate of change of
the scale factor
\begin{equation}
\dot{a}={1\over
\gamma\bar{\mu}}\sin(\bar{\mu}\c)\cos(\bar{\mu}\c).\label{dota}
\end{equation}
Furthermore, the vanishing of the Hamiltonian constraint, ${\cal
H}_{\rm eff}\approx 0$, implies
\begin{equation}
\sin^2(\bar{\mu}\c)={\kappa\gamma^2\bar{\mu}^2\over 3a}{\cal H}_{\rm
M}.\label{sin}
\end{equation}
Combining Eqs. (\ref{dota}) and (\ref{sin}) yields the effective
Friedmann equation for the Hubble rate $H=\dot{a}/a$,
\begin{equation}
H^2={\kappa\over3}\rho\left(1-{\rho\over\rho_{\rm
crit}}\right),\label{Feq}
\end{equation}
with the critical density given by
\begin{equation}
\rho_{\rm crit}={\sqrt{3}\over 16\pi^2\gamma^3}\rho_{\rm
Pl},\label{crit}
\end{equation}
where $\rho_{\rm Pl}=G^{-2}$ is the Planck density. The modified
Friedmann equation provides an effective description for LQC which
very well approximates the underlying discrete quantum dynamics. The
nonperturbative quantum geometric effects are manifested in the
modified Friedmann equation with a $\rho^2$ correction term. The
negative definition of the $\rho^2$ term implies that the Hubble
parameter vanishes when $\rho=\rho_{\rm crit}$ and the universe
experiences a turnaround in the scale factor. When $\rho\ll
\rho_{\rm crit}$, the modifications to the Friedmann equation become
negligible, and the standard Friedmann equation is recovered. It is
remarkable that the origin of $\rho_{\rm crit}$ is purely quantum,
since in the classical limit $\hbar\rightarrow 0$ one has $\rho_{\rm
crit}\rightarrow \infty$. In addition, it should be noted that,
interestingly, $\rho^2$ modifications also appear in string inspired
braneworld scenarios and it has been shown that there exist
interesting dualities between the two frameworks \cite{dualRS}. Such
modifications in braneworlds, however, are usually positive definite
so that a bounce is absent, unless the existence of two timelike
extra dimensions is assumed \cite{bouncingbrane,Piao:2004hr}.

The modified Friedmann equation (\ref{Feq}) along with the
conservation law
\begin{equation}
\dot{\rho}+3H(\rho+p)=0,\label{conservation}
\end{equation}
provides a pre-big-bang picture with a big bounce occurring when
$\rho=\rho_{\rm crit}$. The condition for the existence of a bounce
is that the matter in the universe does not violate the null energy
condition \cite{bounce}. Let us consider the simplest case of matter
with constant equation of state $w$ (the definition of $w$ is
$w=p/\rho$), then from Eq. (\ref{conservation}) we have $\rho\propto
a^{-3(1+w)}$. Thus the conclusion can be drawn that for $w<-1$ a
recollapse can occur and for $w>-1$ a bounce can occur. This can be
easily understood. Matter with equation of state $w<-1$ violates the
null energy condition and hence has increasing energy density as the
universe expands. Thus when the energy density reaches the quantum
critical value a recollapse appears and the universe begins
contracting \cite{Sami:2006wj}. On the other hand, matter with
equation of state $w>-1$ has increasing energy density as the
universe contracts and therefore a quantum bounce will occur when
the quantum critical density is hit. So it is clear that generically
a phantom field can not have a quantum bounce in the early universe
\cite{bounce}. In this paper we will only consider the ``normal''
matter (or canonical scalar field) without null energy condition
violation, and we will focuss on the post-big-bounce stage of the
universe.

\section{Inflationary universe after big bounce}

In this section we shall investigate inflation within the framework
of the effective theory of LQC. In LQC, a quantum bounce plays a
role of junction of pre-big-bang branch and post-big-bang branch, so
the classical big-bang singularity is erased by the quantum
big-bounce. Naturally, the quantum bounce becomes the initial
condition of the subsequent inflation. It is evident that the
quantum gravity effects will play a significant role in the
``primary inflation'', however, what is important is that whether
the quantum geometry effects can leave influences on the
``observable inflation'' with the last 60 $e$-foldings. We shall
discuss some topics of interest of inflation in what follows.

\subsection{Super-inflation and effective quintom}

Though the issue of super-inflation in LQC has been discussed widely
in the literature (see, e.g.,
\cite{dualRS,bounce,suinf1,Tsujikawa:2003vr}), we still make some
analyses here for maintaining this paper self-contained. Besides, we
discuss the effective behavior of a quintom matter in the inflation
process in this subsection.

According to the effective description of the quantum dynamics, the
Friedmann equation is modified to include a $\rho^2$ term relevant
to high energy region, shown as Eq. (\ref{Feq}). Thus, using the
Friedmann equation (\ref{Feq}) and the conservation law
(\ref{conservation}), we have
\begin{equation}
\dot{H}=-{1\over 2}(\rho+p)\left(1-{2\rho\over
p}\right),\label{dotH}
\end{equation}
where $p$ is the pressure of the matter filled in the universe, and
we have set $\kappa=1$ for convenience (this convention will be used
in the most situations hereafter). It is obvious that $\dot{H}$ will
be always larger than zero when the energy density is in the range
of $\rho_{\rm crit}/2<\rho<\rho_{\rm crit}$, provided that
$\rho+p>0$. This implies that a super-inflation phase occurs after
the bounce when the energy is very high. Since $\rho+p>0$ (i.e.
$w>-1$), there is no phantom matter in the universe; so the
super-inflation is purely due to the quantum geometry effects in
LQC.\footnote{It should be noted that super-inflation occurs after
any bounce by definition. Bounce is a point with $H=0$
($\dot{a}=0$), $\ddot{a}>0$, so if we have a bounce, we should have
some period of super-inflation (i.e., growth of $H$ during
expansion) after, independently of a particular theory which causes
this bounce. The bounce discussed in this paper originates from the
effects of LQC, so does the super-inflation.} As shown in the
previous section, matter with $w<-1$ will not lead to big-bounce,
thus in this paper we only consider the matter with $w>-1$ which
does not violate the weak energy condition. From Eqs. (\ref{Feq})
and (\ref{dotH}), we derive
\begin{equation}
{\ddot{a}\over a}=\dot{H}+H^2=-{1\over
6}\left\{\rho\left(1-{\rho\over\rho_{\rm
crit}}\right)+3\left[p\left(1-{2\rho\over \rho_{\rm
crit}}\right)-{\rho^2\over\rho_{\rm
crit}}\right]\right\}.\label{ddota}
\end{equation}
Comparing to the classical form of the equation, it is convenient to
define the effective energy density and pressure
\begin{equation}
\rho_{\rm eff}=\rho\left(1-{\rho\over\rho_{\rm
crit}}\right),\label{rhoeff}
\end{equation}
\begin{equation}
p_{\rm eff}=p\left(1-{2\rho\over\rho_{\rm
crit}}\right)-{\rho^2\over\rho_{\rm crit}},\label{peff}
\end{equation}
then Eq. (\ref{ddota}) can be written as
\begin{equation}
{\ddot{a}\over a}=-{1\over 6}(\rho_{\rm eff}+3p_{\rm eff}).
\end{equation}
Also, we have
\begin{equation}
H^2={1\over 3}\rho_{\rm eff},
\end{equation}
\begin{equation}
\dot{H}=-{1\over 2}(\rho_{\rm eff}+p_{\rm eff}).
\end{equation}
Given the effective energy density and pressure, the effective
equation of state is defined naturally as
\begin{equation}
w_{\rm eff}={p_{\rm eff}\over \rho_{\rm eff}}={w(1-2x)-x\over
1-x},\label{weff}
\end{equation}
where $x$ is defined as dimensionless density, $x=\rho/\rho_{\rm
crit}$, so we have $0<x<1$.

%%%%%%%%%%%%%%%%%%%%%%%%%%%%%%%%%%%%%%%%%%%%%%%%%%%%%%%%%%%%%%%%%%
\begin{figure}[htbp]
\begin{center}
\includegraphics[scale=1.2]{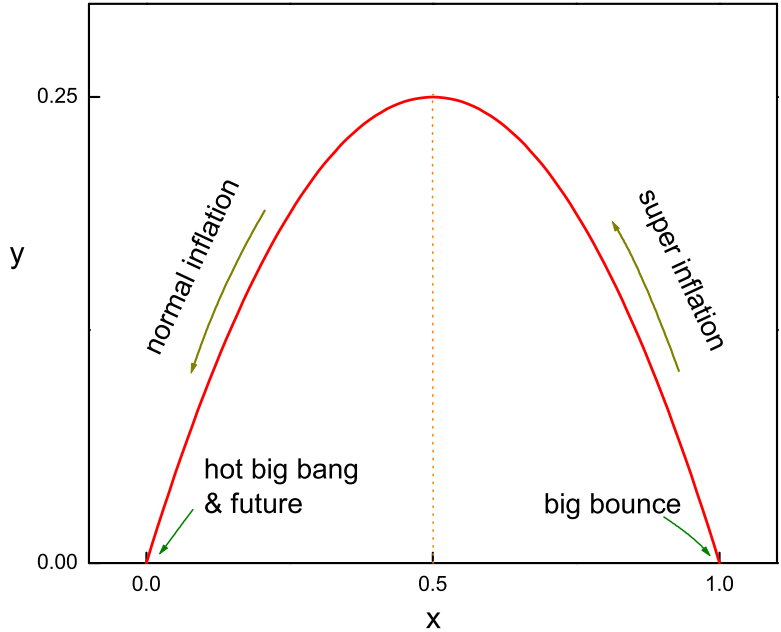}
\caption[]{\small Illustration of inflationary universe in the
effective theory of LQC. Here $y$ denotes $\rho_{\rm eff}/\rho_{\rm
crit}$ and $x$ denotes $\rho/\rho_{\rm crit}$. The plot shows that
after the big-bounce a super-inflation begins due to the quantum
gravity effects during $1/2<x<1$, and subsequently the universe
undergoes a normal-inflation until some reheating process initiates
the ``hot-big-bang''. The dimensionless effective energy density $y$
has a maximum $y_{\rm max}=1/4$ at the end of super-inflation. The
value of $y$ first increases and then decreases, implying an
effective behavior of ``quintom'', which is totally due to the
influence of quantum geometry effects.}\label{fig1}
\end{center}
\end{figure}
%%%%%%%%%%%%%%%%%%%%%%%%%%%%%%%%%%%%%%%%%%%%%%%%%%%%%%%%%%%%%%%%%%%

\begin{figure}[htbp]
\begin{center}
\includegraphics[scale=1.2]{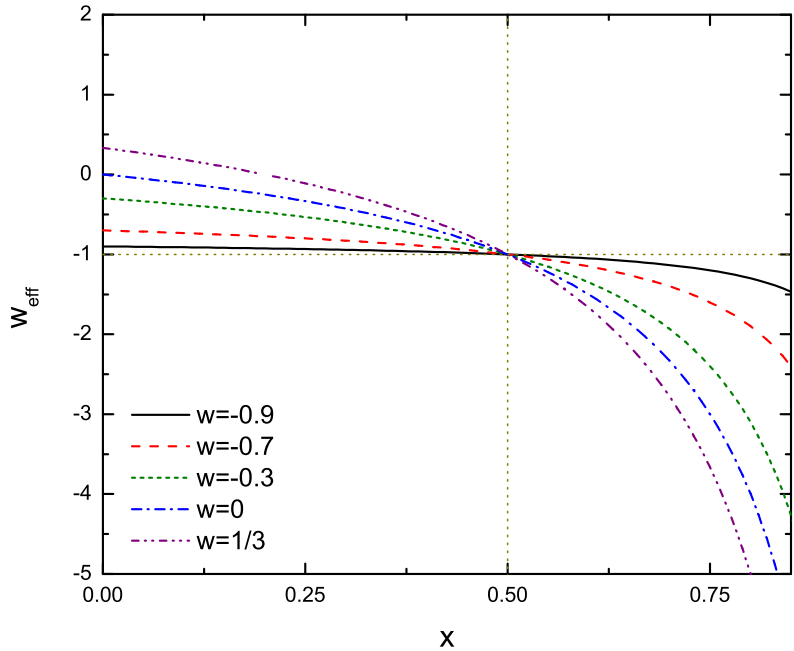}
\caption[]{\small The effective equation-of-state $w_{\rm eff}$
versus the dimensionless density parameter $x$. The
equation-of-state of matter in the universe, $w$, is assumed to be
constant for simplicity, and its value is taken to be $-0.9$,
$-0.7$, $-0.3$, 0 and $1/3$, for example. It is shown that $w_{\rm
eff}$ evolves across the ``phantom divide'' $w=-1$, which is the
significant characteristic of the ``quintom'' matter. Concretely,
$w_{\rm eff}<-1$ when $1/2<x<1$, and $w_{\rm eff}>-1$ when
$0<x<1/2$.}\label{fig2}
\end{center}
\end{figure}

From Eq. (\ref{weff}), we see that provided that the dimensionless
density $x$ is placed in the range $1/2<x<1$, the effective equation
of state will be smaller than $-1$, namely $w_{\rm eff}<-1$, giving
rise to the phase of super-inflation. The super-inflation will last
to $w_{\rm eff}=-1$ when $x=1/2$ and the effective energy density of
the universe $\rho_{\rm eff}$ achieves its maximum value, $\rho_{\rm
crit}/4$. The inflationary process is sketched in Fig. \ref{fig1}
which plots the rewritten Eq. (\ref{rhoeff}), $y=x(1-x)$, where
$y=\rho_{\rm eff}/\rho_{\rm crit}$ and $x=\rho/\rho_{\rm crit}$.
This figure explicitly shows that after the quantum big-bounce of
the universe a super-inflation begins due to the quantum gravity
effects during $1/2<x<1$, and subsequently the universe undergoes a
normal-inflation stage until some reheating process initiates the
``hot-big-bang'' epoch. It is worthwhile to note that the difference
between the conceptions of ``classical big-bang singularity'' and
``hot-big-bang epoch'' should be distinct. The classical big-bang
singularity has been replaced by a quantum bounce in the theory of
LQC, and the hot-big-bang epoch is originated from some reheating
mechanism of the theory of inflation. The hot-big-bang epoch and
future evolution era are expected to happen at the low energy region
(i.e., $x\ll 1$) where the quantum geometry effects can be
negligible. From Fig. \ref{fig1} it can be seen that the value of
$y$ first increases and then decreases, which implies that the
effective behavior of the universe under the quantum gravity
domination resembles a ``quintom'' (for quintom see e.g.
\cite{quintom}). The key feature of quintom is that its equation of
state can evolve across the cosmological-constant boundary (or
``phantom divide'') $w=-1$. The effective equation of state of the
universe in LQC, Eq. (\ref{weff}), just possesses this significant
characteristic, see Fig. \ref{fig2}. In the examples of Fig.
\ref{fig2}, we endow values to the equation of state of matter in
the universe, $w$, as $-0.9$, $-0.7$, $-0.3$, 0 and $1/3$,
respectively. It is shown in this figure that $w_{\rm eff}$ will
evolve across the phantom divide $w=-1$ no matter what matter
component is filled in the universe. Even though the dominant matter
component is dust-like matter ($w=0$) or radiation-like matter
($w=1/3$), the super-inflation and the subsequent normal inflation
will both happen deterministically because of the effective quintom
behavior of the universe due to the quantum gravity nature of the
big-bounce. It is remarkable that recently in Ref.
\cite{quintombounce} the authors pointed out that a bouncing
universe should be filled with quintom matter. Intriguingly, we show
here that an effective quintom behavior is emergent in the bouncing
universe purely due to the effects of quantum geometry in LQC.

%%%%%%%%%%%%%%%%%%%%%%%%%%%%%%%%%%%%%%%%%%%%%%%%%%%%%%%%%%%%%%%%%%
%%%%%%%%%%%%%%%%%%%%%%%%%%%%%%%%%%%%%%%%%%%%%%%%%%%%%%%%%%%%%%%%%%%

\subsection{Condition for inflation}

In this subsection we study the condition of inflation in detail
within the framework of LQC. In the previous subsection we see that
after the big-bounce a super-inflation is emergent naturally and a
normal inflation occurs subsequently. The emergence of the
super-inflation is purely a phenomena of quantum gravity, which is
irrelevant to what matter is filled in the universe. The subsequent
normal inflation will also happen deterministically, but whether it
can last sufficient $e$-foldings is relevant to the matter
component.

\begin{figure}[htbp]
\begin{center}
\includegraphics[scale=1.2]{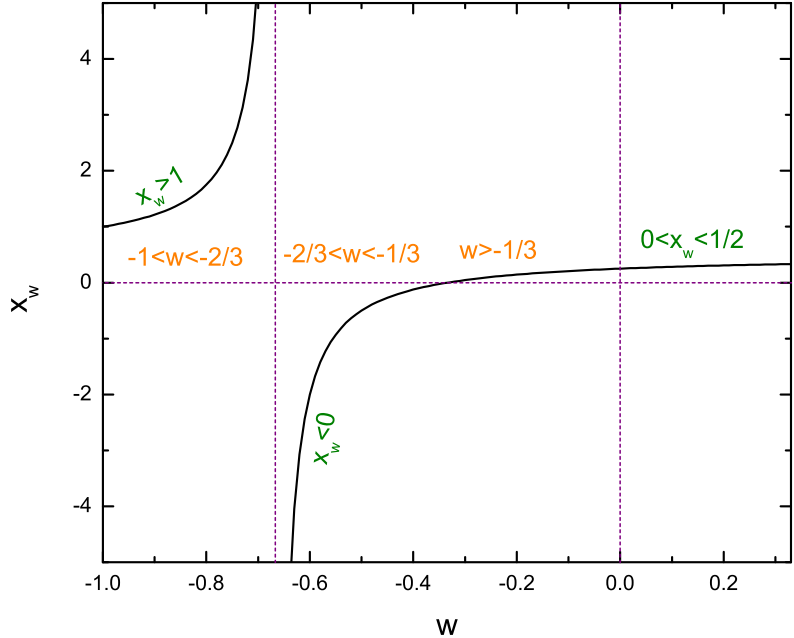}
\caption[]{\small The plot of $x_w$ versus $w$. It is shown that
$0<x_w<1/2$ for $w>-1/3$, $x_w<0$ for $-2/3<w<-1/3$, and $x_w>1$ for
$-1<w<-2/3$. }\label{fig3}
\end{center}
\end{figure}

Generically, the criteria of judgement for inflation is attributed
to $\epsilon_H<1$, where the parameter $\epsilon_H$ is defined as
$\epsilon_H=-\dot{H}/H^2$. Using Eqs. (\ref{Feq}) and (\ref{dotH}),
we can easily derive
\begin{equation}
\epsilon_H={3\over 2}(1+w){1-2x\over 1-x}.\label{epsilonH}
\end{equation}
Evidently, the condition $\epsilon_H<1$ is totally equivalent to the
condition $w_{\rm eff}<-1/3$. Solving the inequality $\epsilon_H<1$
or $w_{\rm eff}<-1/3$ yields a set of two inequalities: $x>x_w$ for
$w>-2/3$, $x<x_w$ for $w<-2/3$, where $x_w$ is defined as
\begin{equation}
x_w={1+3w\over 4+6w}.\label{xw}
\end{equation}
However, since $0<x<1$, we have to analyze whether or not $x_w$ is
located in the range $x\in (0,~1)$. We find from Eq. (\ref{xw}) that
$0<x_w<1/2$ for $w>-1/3$, $x_w<0$ for $-2/3<w<-1/3$, and $x_w>1$ for
$-1<w<-2/3$. For sketching the cases of $x_w$ values clearly, we
plot Eq. (\ref{xw}) in Fig. \ref{fig3}.

%%%%%%%%%%%%%%%%%%%%%%%%%%%%%%%%%%%%%%%%%%%%%%%%%%%%%%%%%%%%%%%%%%
%%%%%%%%%%%%%%%%%%%%%%%%%%%%%%%%%%%%%%%%%%%%%%%%%%%%%%%%%%%%%%%%%%%

Given $0<x<1$ as well as the analysis of $x_w$, the condition of
$x>x_w$ for $w>-2/3$ is divided into two parts: $x>x_w$ for
$w>-1/3$, and $x>0$ for $-2/3<w<-1/3$; the condition of $x<x_w$ for
$-1<w<-2/3$ becomes $x<1$ for $-1<w<-2/3$. Therefore, we have the
condition for inflation as follows

%\begin{equation} \label{cond}
%\left\{ \begin{aligned}
 %        0<x_w<x<1, ~{\rm for}~w>-1/3;\\
 %                 0<x<1,~{\rm for}~-1<w<-1/3.
  %                        \end{aligned} \right.
  %                        \end{equation}

\begin{equation} \label{cond}
\begin{split}
0<x_w<x<1, ~{\rm for}~w>-1/3; \\
0<x<1, ~{\rm for}~-1<w<-1/3.
 \end{split}
 \end{equation}
It is obvious that when $-1<w<-1/3$ inflation will always happen
which is in accordance with the usual cases of inflation. However,
when $w>-1/3$, inflation will also take place for period of time,
which is different from the usual inflation cases and is thus purely
due to the influence of the quantum geometry. In particular, the
inflation process is divided into two stages, super-inflation and
normal-inflation. For super-inflation, the condition is
\begin{equation}
1/2<x<1,~ {\rm for~any~} w>-1,
\end{equation}
which has been discussed in the previous subsection. For normal
inflation, the condition is
%\begin{equation} \label{cond2}
%\left\{ \begin{aligned}
%         0<x_w<x<1/2, ~{\rm for}~w>-1/3~;\\
 %                 0<x<1/2,~{\rm for}~-1<w<-1/3~.
 %                         \end{aligned} \right.
 %                         \end{equation}
\begin{equation} \label{cond2}
\begin{split}
0<x_w<x<1/2, ~{\rm for}~w>-1/3; \\
0<x<1/2, ~{\rm for}~-1<w<-1/3.
 \end{split}
 \end{equation}

So far we see that the quantum geometry effects lead to intriguing
phenomena in the inflationary stage after the big-bounce in LQC.
First, any matter with $w>-1$ can give rise to super-inflation after
the bounce when $1/2<x<1$. In addition, matter with $w>-1/3$ can
proceed to drive the subsequent normal inflation for period of time.
So, even though the universe is filled in radiation-like or
dust-like component at that time, inflation will still happen. These
peculiar phenomena all root in the effects of quantum geometry in
LQC. A question is now emergent: How many $e$-foldings can this LQC
induced inflation last?

\subsection{$e$-foldings}

We now discuss the issue of $e$-folding number in this subsection.
It is of interest to know how many $e$-foldings the inflation can
last when $w>-1/3$ in the framework of the effective dynamics of
LQC. Also, for $-1<w<-1/3$, it is necessary to evaluate when the
inflation should cease given the number of $e$-foldings.

For simplicity we assume that the equation of state of the matter in
the universe is a constant. Such an assumption makes Eq.
(\ref{conservation}) have a simple solution,
\begin{equation}
\rho=\rho_{\rm crit}\left({a\over a_{\rm crit}}\right)^{-3(1+w)},
\end{equation}
where $a_{\rm crit}$ denotes the scale factor at bouncing point when
the energy density approaches the critical value, so $a_{\rm crit}$
is the minimum scale for the universe, $a>a_{\rm crit}$. According
to the definition of the parameter $x$, $x=\rho/\rho_{\rm crit}$, we
have $x=({a/ a_{\rm crit}})^{-3(1+w)}$, then we obtain
\begin{equation}
e^{{\cal N}}\equiv {a\over a_{\rm crit}}=x^{-{1\over 3(1+w)}},
\end{equation}
where ${\cal N}$ is the number of $e$-foldings from the bouncing
point (corresponding to $x=1$) to somewhere in inflation stage
(labeled by $x$).

%%%%%%%%%%%%%%%%%%%%%%%%%%%%%%%%%%%%%%%%%%%%%%%%%%%%%%%%%%%%%%%%%%
\begin{figure}[htbp]
\begin{center}
\includegraphics[scale=1.2]{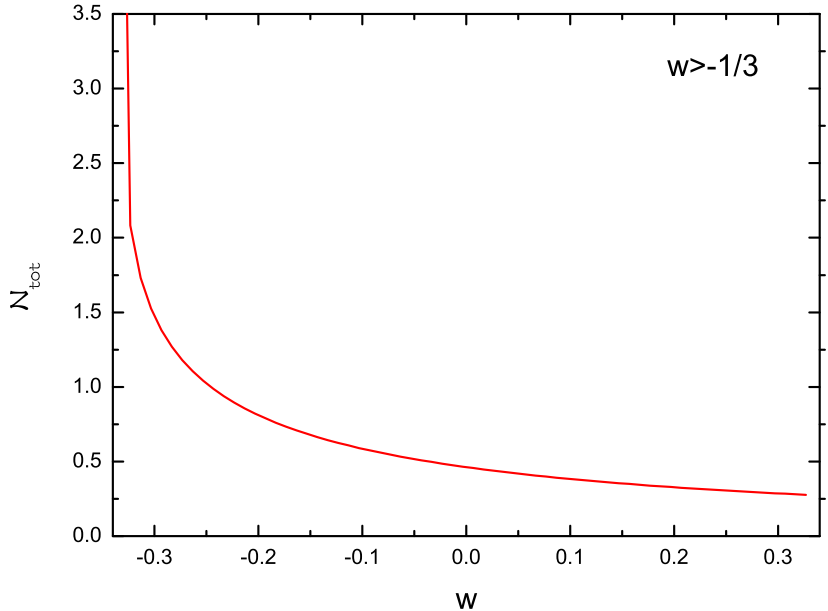}
\caption[]{\small The relation between the total $e$-folding number
${\cal N}_{\rm tot}$ and the equation-of-state $w$, when $w>-1/3$.
The inflation driven by such an equation of state originates from
the quantum geometry effects in LQC. The plot shows that such an
inflation can last at most several $e$-foldings.}\label{fig4}
\end{center}
\end{figure}
%%%%%%%%%%%%%%%%%%%%%%%%%%%%%%%%%%%%%%%%%%%%%%%%%%%%%%%%%%%%%%%%%%%

Therefore, for cases of $w>-1/3$, the total number of $e$-foldings
can be evaluated directly,
\begin{equation}
{\cal N}_{\rm tot}=-{1\over 3(1+w)}\ln x_w,
\end{equation}
where ${\cal N}_{\rm tot}$ represents the total $e$-folding number,
and $x_w$ is given by Eq. (\ref{xw}). The inflation driven by such
an equation of state will terminate at $x_w$, so the total
$e$-folding number is a function of $w$, namely ${\cal N}_{\rm
tot}(w)$, see Fig. \ref{fig4}. This figure shows that such an
inflation can last at most several $e$-foldings, e.g., ${\cal
N}_{\rm tot}(1/3)=0.275$, ${\cal N}_{\rm tot}(0)=0.462$, ${\cal
N}_{\rm tot}(-0.30)=1.472$, and ${\cal N}_{\rm tot}(-0.33)=2.641$.

For cases of $-1<w<-1/3$, inflation can always proceed, thus a
reheating mechanism is needed to cease the inflation and initiate
the hot-big-bang era. The concrete mechanism of reheating is not a
issue of concern in this paper. What is of interest for us is that
one can evaluate the value of $x$ when the inflation ends if $w$ and
${\cal N}_{\rm tot}$ are given. The end point of inflation can be
expressed as
\begin{equation}
x_{\rm end}=e^{-3(1+w){\cal N}_{\rm tot}}.
\end{equation}
In fact, the total number of $e$-folds of inflation is an unknown
value. Our observational universe corresponds to roughly last 60
$e$-folds before the end of inflation, hence at most the last 60
$e$-folds have a directly observational effect. Inflation stage
before the observational universe leaves the horizon can not be
observed in principle, so this stage is called ``primary
inflation''. As oppose to this unobservable stage, we call the
inflation stage corresponding to the observational universe the
``observational inflation''. Hence, the total number of $e$-foldings
of inflation must be larger than 60, namely ${\cal N}_{\rm tot}>60$,
and may even reach several hundreds or more. Note that the value of
$x_{\rm end}$ is related to the energy scale of reheating since
$x_{\rm end}=\rho_{\rm end}/\rho_{\rm crit}$.

\subsection{Slow-roll inflation and primordial perturbations}

During inflation the wavelengths of perturbations generated from
short scales where quantum gravity effects are important are
stretched to cosmic scales by rapid expansion, thus the effects of
quantum gravity should in principle leave an imprint on the
primordial spectrum of perturbations. Quantum geometry effects in
LQC lead to a $\rho^2$ modification in the Friedmann equation which
gives rise to a quantum bouncing solution to replace the classical
big-bang singularity in the very early universe. The quantum gravity
effects, however, only play a significant role in the very high
energy regime, the quantum correction $\rho^2$ term will be
negligible when the energy scale is much smaller than the scale of
the critical density. Still, it is expected that the imprint of
quantum effects should be left on the primordial power spectrum
which can be investigated through observation on the CMB sky. It
should be mentioned that the stringy imprint in the primordial
perturbations has been discussed in the noncommutative inflation
models, see e.g., \cite{ncinfl}.

While cosmological scales are leaving the horizon, the slow-roll
paradigm of inflation is practically mandatory in order to account
for the near scale invariance of spectrum of the primordial
curvature perturbation. We study the slow-roll inflation within the
framework of the effective theory of LQC, i.e., the Friedmann
equation given by Eq. (\ref{Feq}). The inflation process is driven
by a spatially homogeneous scalar field $\phi$ (the inflaton)
satisfying the following equation of motion,
\begin{equation}
\ddot{\phi}+3H\dot{\phi}+V'(\phi)=0,\label{eom}
\end{equation}
where the prime denotes the derivative with respect to the field
$\phi$, and the Hubble parameter $H$ is given by Eq. (\ref{Feq}). If
$\dot{\phi}^2\ll V(\phi)$ and $\ddot{\phi}\ll 3H\dot{\phi}$, the
scalar field will slowly roll down its potential, and the exact
evolution equation (\ref{eom}) can be replaced by the slow-roll
approximation,
\begin{equation}
\dot{\phi}=-V'/3H.\label{slow}
\end{equation}
Under the slow-roll condition, the energy density of the scalar
field approximates as $\rho\sim V(\phi)$, the Friedmann equation
(\ref{Feq}) consequently can be written as
\begin{equation}
H^2={1\over 3} V\left(1-{V\over \rho_{\rm crit}}\right).\label{Feq2}
\end{equation}
For convenience one can define an effective potential for the scalar
field,
\begin{equation}
V_{\rm eff}(\phi)=V(\phi)(1-\nu(\phi)),
\end{equation}
where $\nu(\phi)$ is defined as
\begin{equation}
\nu(\phi)={V(\phi)\over \rho_{\rm crit}}.\label{nu}
\end{equation}
It is obvious that $\nu(\phi)$ describes the quantum geometry
effects in LQC. The slow-roll parameters can be defined in terms of
the potential and its derivatives as usual,
\begin{equation}
\epsilon_v={1\over 2}\left({V'\over V}\right)^2, ~~~~
\eta_v={V''\over V},~~~~ \xi_v^2={V'V'''\over V^2},
\end{equation}
then the slow-roll condition can be expressed as $\epsilon_v,
~|\eta_v|\ll 1$. The inflation ends when the slow-roll condition
ceases to be satisfied. The small change of $e$-foldings satisfies
$dN\equiv -Hdt$; using Eqs. (\ref{slow}) and (\ref{Feq2}), one
obtains the number of $e$-foldings of slow-roll inflation remaining
at a given epoch,
\begin{equation}
N(\phi)=\int\limits_{\phi_{\rm end}}^{\phi}{V_{\rm eff}\over
V'}d\phi=\int\limits_{\phi_{\rm end}}^{\phi}(1-\nu){V\over
V'}d\phi,\label{efold}
\end{equation}
where $\phi_{\rm end}$ marks the end of slow-roll inflation.

The perturbation $\delta\phi$ can be treated as a massless free
field, and its vacuum fluctuation can be regarded as a classical
quantity a few Hubble times after horizon exit. The spectrum of the
field perturbation is
\begin{equation}
{\cal P}_\phi=(H/2\pi)^2.\label{Pphi}
\end{equation}
The corresponding curvature perturbation is given by ${\cal
R}=(H/\dot{\phi})\delta\phi$. Using Eqs. (\ref{slow}), (\ref{Feq2})
and (\ref{Pphi}), we get the amplitude of scalar perturbation as
\begin{equation}
{\cal P_R}={H^2\over {\dot{\phi}}^2}\left({H\over
2\pi}\right)^2={V_{\rm eff}^3\over 12\pi^2 V'^2}={V^3\over 12\pi^2
V'^2}(1-\nu)^3,\label{PR}
\end{equation}
which is evaluated at the Hubble radius crossing $k=aH$. Since $H$
is slowly varying, we have $d\ln k=d(\ln (aH))\simeq d\ln a=Hdt$,
then from Eqs. (\ref{slow}) and (\ref{Feq2}), we get
\begin{equation}
{d\over d\ln k}=-{V'\over V_{\rm eff}}{d\over d\phi}=-{1\over
1-\nu}{V'\over V}{d\over d\phi}.
\end{equation}
We shall need the following expressions:
\begin{equation}
{d\epsilon_v\over d\ln k}={1\over
1-\nu}(-2\epsilon_v\eta_v+4\epsilon_v^2),
\end{equation}
\begin{equation}
{d\eta_v\over d\ln k}={1\over 1-\nu}(2\epsilon_v\eta_v-\xi_v^2).
\end{equation}
The spectral index of the scalar perturbation and its running can
thus be given,
\begin{equation}
n_s-1\equiv{d\ln {\cal P_R}\over d \ln k}=-6\epsilon_v{1-2\nu\over
(1-\nu)^2}+2\eta_v{1\over 1-\nu},\label{ns}
\end{equation}
\begin{equation}
\alpha_s\equiv {dn_s\over d\ln k}=16\epsilon_v\eta_v{1-2\nu\over
(1-\nu)^3}-24\epsilon_v^2{1-3\nu(1-\nu)\over
(1-\nu)^4}-2\xi_v^2{1\over (1-\nu)^2}.\label{running}
\end{equation}
Inflation also generates gravitational waves with two independent
components $h_{+,\times}$ which have the same action as a massless
scalar field. Likewise, the amplitude of tensor perturbations can
also be given,
\begin{equation}
{\cal P}_{\rm grav}=8\left({H\over 2\pi}\right)^2={2V_{\rm eff}\over
3\pi^2}={2 V\over 3\pi^2}(1-\nu),
\end{equation}
which is also evaluated at the horizon exit $k=aH$. Then we obtain
the spectral index of tensor perturbation and the scalar-to-tensor
ratio,
\begin{equation}
n_{\rm grav}\equiv {d\ln {\cal P}_{\rm grav}\over d\ln
k}=-2\epsilon_v{1-2\nu\over (1-\nu)^2},\label{nT}
\end{equation}
\begin{equation}
r\equiv {{\cal P}_{\rm grav}\over {\cal P_R}}=16\epsilon_v{1\over
(1-\nu)^2}.\label{r}
\end{equation}
Hence, the consistency relation can be expressed as
\begin{equation}
r=-8 n_{\rm grav}{1\over 1-2\nu}.
\end{equation}
Evidently, all quantities will recover the standard forms of
slow-roll inflation when $\nu$ approaches zero. In addition, since
$\nu$ is certainly a small quantity, we thus have $n_s-1\approx
-6\epsilon_v+2\eta_v(1+\nu)$, $\alpha_s\approx
16\epsilon_v\eta_v(1+\nu)-24\epsilon_v^2(1+\nu)-2\xi_v^2(1+2\nu)$,
$r\approx 16\epsilon_v(1+2\nu)$, and so on.

For concreteness, we consider a simple inflation model, chaotic
inflation, for example. The chaotic inflation has the potential of
the form $V(\phi)\propto\phi^\alpha$ with $\alpha$ a positive
integer. The slow-roll parameters can be easily expressed as
$\epsilon_v=\alpha^2/2\phi^2$ and
$\eta_v=2(\alpha-1)\epsilon_v/\alpha$. Strictly speaking, the
termination of the inflation should be determined by the condition
$\epsilon_H\simeq 1$, where $\epsilon_H\sim
\epsilon_v(1-2\nu)/(1-\nu)$. Nevertheless, the fact that $\nu$ is a
small quantity implies $\epsilon_N\simeq \epsilon_v$ showing the
usual termination condition $\epsilon_v\simeq 1$ is also
appropriate. In any case, we have $\phi_{\rm end}\ll \phi_N$, where
$\phi_N$ (with $N\sim 60$) marks the value of the field when our
observable universe leaves the horizon during the inflation. Using
Eq. (\ref{efold}) and integrating out $N$, we derive
\begin{equation}
\phi_N=\sqrt{{2\alpha N\over (1-{2\over \alpha+2}\nu)}}.
\end{equation}
Consequently, the concrete expressions of the slow-roll parameters
can be obtained,
\begin{equation}
\epsilon_v={\alpha\over 4N}\left(1-{2\over \alpha+2}\nu\right),
\end{equation}
\begin{equation}
\eta_v={\alpha-1\over 2N}\left(1-{2\over \alpha+2}\nu\right),
\end{equation}
\begin{equation}
\xi_v^2={(\alpha-1)(\alpha-2)\over 4 N^2}\left(1-{2\over
\alpha+2}\nu\right)^2.
\end{equation}
Then we get all quantities of interest in cosmology, such as scalar
spectral index and its running as well as the tensor-to-scalar
ratio, using Eqs. (\ref{ns}), (\ref{running}) and (\ref{r}).

\subsection{Can LQC effects be observable?}

Since inflation manifests fluctuations that were once on scales of
quantum gravity dominance to scales of the observable horizon,
quantum gravity physics could potentially leave its imprint on the
CMB sky. It is of interest to discuss the possibility of detecting
the signature of LQC physics in the angular power spectrum.
Discussions in the previous subsection have revealed the possible
forms of the primordial power spectrum and other observational
quantities of interest. The expected amplitude of quantum gravity
effects can be characterized by a parameter $\nu$ which appears in
the resulting quantities in cosmology from slow-roll inflation.
Naively, one can determine the value of $\nu$ by fitting the
theoretical results to observational data.

Actually, the amplitude of $\nu$ can easily be constrained using the
information of the COBE normalization \cite{cobe4}. The COBE
normalization corresponds to $\delta_H=(2/5){\cal
P_R}^{1/2}=1.91\times 10^{-5}$ for the mode which crossed the Hubble
radius about 60 $e$-folds before the end of inflation. Comparing
this value to Eq. (\ref{PR}) gives
\begin{equation}
{V^{1/4}(1-\nu)^{3/4}\over \epsilon_v^{1/4}}=2.7\times 10^{-2}
M_{\rm Pl}=6.7\times 10^{16}~{\rm GeV},\label{cobe}
\end{equation}
where $M_{\rm Pl}=2.4\times 10^{18}~{\rm GeV}$ has been used. The
value of the critical density in the effective theory of LQC can be
deduced from Eq. (\ref{crit}) as $\rho_{\rm crit}\approx
0.82\rho_{\rm Pl}$. The Planck density $\rho_{\rm
Pl}=G^{-2}=2.22\times 10^{76}~{\rm GeV}^4$, so we have $\rho_{\rm
crit}^{1/4}=1.162\times 10^{19}~{\rm GeV}$. Using Eqs. (\ref{nu})
and (\ref{cobe}), and in view of $\nu$ is a small quantity, we get
\begin{equation}
\nu\simeq 10^{-9}\epsilon_v.
\end{equation}
Since $\epsilon_v$ is much less than 1, the LQC parameter $\nu$ is
certainly much smaller than $10^{-9}$. Therefore, we conclude that
the loop quantum effects can only lead to a very tiny imprint in the
primordial power spectrum and their signature can hardly be detected
in the CMB sky using present and upcoming observational data.

%\section{Discussions}

\section{Conclusions}

In this paper we have investigated the inflationary universe in the
framework of the effective theory of LQC. In LQC, the
nonperturbative quantum geometry effects lead to a $\rho^2$ term
with a negative sign in the modified Friedmann equation. This
semiclassical theory gives rise to a quantum bounce in the high
energy regime when the loop quantum effects are dominative power,
which is in accordance with the result from the quantum evolution in
LQC. The classical big-bang singularity is thus replaced by the
quantum bounce. After the bounce, a super-inflation phase was
emergent in a natural way. Then the universe underwent a normal
inflation stage.

No matter what matter component (even if radiation-like or dust-like
matter) dominated the universe in the early era after the bounce,
the super-inflation and the subsequent normal inflation would both
happen deterministically, since the quantum gravity dynamics led to
effective quintom behavior for the universe. The effective equation
of state $w_{\rm eff}$ crossed the phantom divide (i.e., $w_{\rm
eff}$ evolved from $<-1$ to $>-1$) in the LQG epoch. The
super-inflation took place in the range of $1/2<x<1$ and the
subsequent normal inflation happened in $0<x<1/2$. It has been shown
that matter with $w>-1/3$ can proceed to drive the normal inflation
for period of time, though it can last at most several $e$-folds.
This peculiar behavior also manifests the quantum geometry effects
in LQC. For $-1<w<-1/3$, the inflation would always take place, thus
a reheating mechanism is needed for terminating the inflation.

Since the cosmological perturbations might be generated when the
universe was in the quantum gravity regime, the LQG physics could
potentially leave its imprint on the CMB sky. We studied the
slow-roll inflation using the semiclassical theory of LQC, namely
the modified Friedmann equation inspired from LQC. Taking the LQC
effects into account, the power spectra of curvature perturbations
and gravitational waves have been derived. We showed that the scalar
spectral index, its running and the tensor-to-scalar ratio can be
expressed in terms of slow-roll parameters as well as the LQC
parameter $\nu$. Note that we analyzed the slow-roll inflation only
using the modified Friedmann equation derived from LQC, ignoring
some LQG corrections in the perturbation equations. More
sophisticated analysis on the cosmological perturbation theory with
LQG corrections see Refs. \cite{pert1,pert2}. The main effects of
LQC have been, however, involved in our results. By analyzing the
power spectrum of the curvature perturbations with the information
of the COBE normalization, we showed that the imprint of the loop
quantum effects is too weak to be observed by present and upcoming
observational data.

\acknowledgments

We would like to thank Rong-Gen Cai, Miao Li, Yun-Song Piao, and Yi
Zhang for helpful discussions. This work is supported by the China
Postdoctoral Science Foundation, the K. C. Wong Education Foundation
(Hong Kong), and the Natural Science Foundation of China. XZ
acknowledges the hospitality of the Center for Gravity and
Relativistic Astrophysics during his visit to Nanchang University.

%%%%%%%%%%%%%%%%%%%%%%%%%%%%%%%%%%%%%%%

\end{document}